\documentclass[twocolumn,showpacs,preprintnumbers,amsmath,amssymb]{revtex4}
\usepackage{graphicx}
\usepackage{dcolumn}
\usepackage{bm}

\begin{document}
\title{A "saddle-node" bifurcation scenario for birth or destruction
of a Smale--Williams solenoid}
\author{Olga~B.~Isaeva,$^1$ Sergey~P.~Kuznetsov,$^{1,2}$
and Igor~R.~Sataev$^1$}
\affiliation{$^1$Kotel'nikov's Institute of Radio-Engineering 
and Electronics of RAS, Saratov Branch, \\ Zelenaya 38, Saratov, 410019, Russian Federation
\\$^2$Institute of Physics and Astronomy, University of Potsdam, \\
Karl-Liebknecht-Str. 24/25, 14476 Potsdam-Golm, Germany}

\date{\today}

\begin{abstract}
Formation or destruction of hyperbolic chaotic attractor under parameter 
variation is considered with an example represented by Smale--Williams 
solenoid in stroboscopic Poincar\'{e} map of two alternately excited 
non-autonomous van der Pol oscillators. The transition occupies a narrow 
but finite parameter interval and progresses in such way that periodic 
orbits constituting a "skeleton" of the attractor undergo saddle-node 
bifurcation events involving partner orbits from the attractor and from a 
non-attracting invariant set, which forms together with its stable manifold 
a basin boundary of the attractor.
\end{abstract}


\maketitle

\section{Introduction}

Uniformly hyperbolic strange attractors have strong chaotic properties and 
are structurally stable, i.e. insensitive to variations of functions and 
parameters in the dynamical equations~\cite{1,2,3,4,5,6,7,8,9}. Formal 
examples are Smale--Williams solenoid, Plykin attractor, and some other 
mathematical constructions suggested mainly in 1960th--1970th~\cite{3,4,5,6}. 
Recently a number of physically realizable systems with hyperbolic 
attractors were proposed~\cite{10,11,12,13}.

One promising direction of search for real-world nonlinear dissipative systems 
with the hyperbolic chaotic attractors may be based on consideration of 
scenarios of their appearance under variation of control parameters. (Note 
that this is an issue certainly different from the commonly known scenarios 
of transition to chaos, like that of Feigenbaum, which lead to non-hyperbolic 
attractors.) Possible appearance of hyperbolic chaotic attractors was discussed 
by Ruelle and Takens in the context of the onset of turbulence~\cite{14,15}, but 
they did not consider concrete examples of dynamical equations demonstrating the 
phenomenon. More recently, Shilnikov and Turaev~\cite{16,17} have suggested a 
scenario of emergence of attractor of Smale--Williams type in a kind of the 
so-called blue sky catastrophe. An example of dynamical equations manifesting 
this scenario was presented and studied numerically in Ref.~\cite{18}.

In this paper we discuss an alternative scenario for the onset or destruction 
of attractor of Smale--Williams type associated with collision of two chaotic 
invariant sets, an attracting and a non-attracting one. Concretely, we consider 
the phenomenon in a physically realizable system composed of two alternately 
activated non-autonomous self-oscillators, which pass the excitation each other 
in such way that the phases of oscillations at successive stages of activity 
evolve chaotically in accordance with an expanding circle map (Bernoulli 
map)~\cite{10,19}. In this system attractor of Smale--Williams type occurs in 
the stroboscopic Poincar\'{e} map. At certain parameters the uniformly hyperbolic 
nature of this attractor was verified in computations~\cite{10,12,19,20}. Here we 
consider birth or disappearance of the chaotic attractor under variation of a 
parameter controlling a relative duration of the stages of activity and silence. 
The scenario may be thought of as a multitude of pairwise collisions of orbits 
relating to the attractor with those from some non-attracting invariant set 
happening in a narrow but finite parameter interval. In Appendix we shortly 
discuss a transparent example of a simple model map demonstrating qualitatively 
the same scenario of birth and destruction of chaotic attractor.

\section{The basic model}

The system we will study is governed by a set of differential 
equations~\cite{10,19}
\begin{equation} \label{1} 
\begin{array}{l} 
\dot{x}=\omega_0 u, \\ 
\dot{u}=(h+A\cos\frac{2\pi t}{T}-x^2)u-\omega_0 x+\frac{\varepsilon}{\omega_0}y\cos\omega_0 t, \\ 
\dot{y}=2\omega_0v, \\ 
\dot{v}=(h-A\cos\frac{2\pi t}{T}-y^2)v-2\omega_0 y+\frac{\varepsilon}{2\omega_0}x^2,
\end{array} 
\end{equation} 
slightly modified by introducing an additional parameter~$h$. The model is 
composed of two subsystems, the van der Pol oscillators with characteristic 
frequencies~$\omega_0$ and $2\omega_0$. The dynamical variables $x$ and $y$ are 
the generalized coordinates of the oscillators while $u$ and $v$ are the 
normalized velocities. In each oscillator the parameter responsible for the birth 
of the limit cycle is forced to vary slowly with period $T$ and amplitude $A$, in 
opposite phases for the two subsystems, which become active turn by turn. The 
newly introduced parameter $h$ controls the relative duration of the stages of 
activity and damping. The coupling between the subsystems characterized by 
coefficient $\varepsilon$ is established in such special way that the excitation 
transfer between the subsystems is accompanied by doubling of the phase shift 
attributed to the oscillations on each next cycle of the transfer. Due to this, 
the stroboscopic map of the system~\eqref{1} in a wide parameter range gives rise 
to a hyperbolic chaotic attractor of Smale--Williams type as it was verified in 
computations based on the cone criterion~\cite{19,12} and on the statistics of 
angles between the stable and unstable manifolds~\cite{10,12}. At a particular 
parameter set $h=0$, $\omega_0=2\pi$, $T=6$, $A=5$, $\epsilon=0.5$ the uniformly 
hyperbolic nature of the attractor was confirmed with a mathematically rigorous 
computer-assisted proof~\cite{20}.

\section{Numerical results and discussion}

\begin{figure}
\includegraphics[width=0.45\textwidth,keepaspectratio]{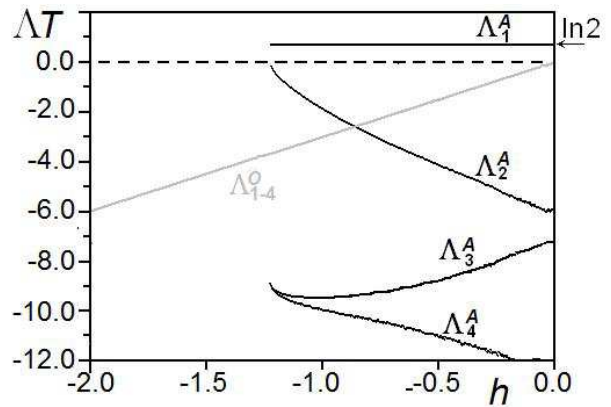}
\caption{The Lyapunov exponents of the system~\eqref{1} versus parameter~$h$ 
(black curves) and the Lyapunov exponent for the trivial attractor, the fixed 
point at the origin (gray line). Parameter values are $\omega_0=2\pi$, $T=6$, 
$\varepsilon=0.5$.}\label{f1}
\end{figure}

Let us start from the case $h=0$, where the operation mode corresponding to the 
Smale--Williams attractor certainly takes place, and move to the domain $h<0$ 
increasing the absolute value of $h$. Then, duration of the activity stages 
will decrease, and at some place the transferred excitation level becomes 
insufficient to recover the amplitude of the oscillations. Then, the system 
drops after some transient to the trivial stable state of zero amplitude. 
Figure~\ref{f1} shows the Lyapunov exponents of the Poincar\'{e} map versus parameter~$h$. 
The right-hand part of the plot corresponds to situation of existence of chaotic 
uniformly hyperbolic attractor of Smale--Williams type. A characteristic feature 
is that the largest Lyapunov exponent is positive, close to $\ln 2\approx 0.693$ 
while others are negative, smoothly depending on the parameter. In fact, at $h<0$ 
this attractor coexists with another one, the trivial stable zero state. (It is 
so because the coupling term in the last equation~\eqref{1} has quadratic 
dependence on the dynamical variable; thus, with initial amplitude small enough, 
the system inevitably relaxes to zero state.) As seen from Fig.~\ref{f1}, with decrease 
of $h$, the breakdown of the chaotic attractor occurs at certain parameter value. As the 
remaining attractor is the fixed point in the origin, after the breakup of the 
chaotic attractor all trajectories tend there. 

The sudden destruction of the chaotic attractor indicates the crises-like 
("dangerous") character of the transition and makes one to suspect that the 
event is associated with a collision of two invariant sets (cf.~\cite{21}). 
To bring a qualitative basis for this observation let us develop some 
approximate approach to stroboscopic description of the dynamics. It is 
convenient to introduce polar coordinates on the plane of the dynamical 
variables of the first oscillator. We set $\{x_0,x_1\}=\{x,u/0.9-x/2\}$ 
and introduce the amplitude (or radial component) and the phase (or angular 
coordinate) in such way that $x_0+ix_1=re^{i\varphi}$. (In definition of the 
variable change the coefficients are chosen to get the form of the attractor 
close to a circular one in the plane projection.) Now, we construct a 
one-dimensional amplitude map for the first oscillator as follows. Starting 
with zero amplitude of the second oscillator and assigning some amplitude~$r$ 
and phase~$\varphi$ to the first oscillator, we evaluate the amplitude~$r_{\rm{new}}$ 
after one period of the parameter modulation and plot it as a function of the 
initial amplitude. The constructed map is not so bad tool for sketchy description 
of the stroboscopic amplitude dynamics: since the subsystems are excited 
alternately, each epoch of activity for one of them corresponds to silence 
period for another one, whose amplitude in the rough approximation can be 
neglected there. 

\begin{figure}
\includegraphics[width=0.45\textwidth,keepaspectratio]{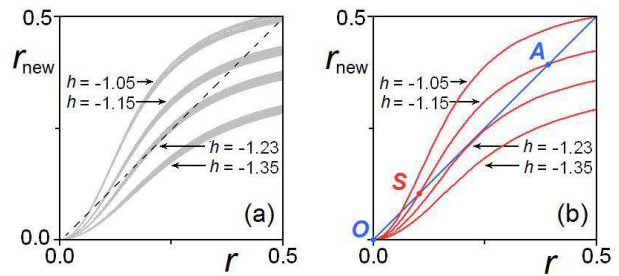}
\caption{Constructing the amplitude stroboscopic map at $\omega_0=2\pi$, $A=6.5$, 
$T=6$, $\varepsilon=0.5$ for different values of parameter $h$: (a)~the original 
diagram accounting all accumulated computational data, and (b)~the idealized 
plot excluding the widening by the phase averaging.}\label{f2}
\end{figure}

Figure~\ref{f2} shows the plots for the amplitude map at four different values of~$h$ 
as obtained from numerical integration of equations~\eqref{1}. Actually, instead 
of curves we get the widened strips on the panel~(a); the reason is that the 
positions of the plotted points depend not only on amplitudes but also to some 
extent on the initial phases, varied from $0$ to $2\pi$. Roughly, this widening 
measures a degree of incorrectness of the description in terms of the amplitude 
map. On the panel~(b) this widening is excluded by means of averaging over 
the initial phases.

As seen from Fig.~\ref{f2}b, with variation of~$h$ a tangent bifurcation occurs. 
At larger~$h$ the map possesses three fixed points (on the plot they 
correspond to crossings of the curve with the bisector). One is located 
in the origin that is the stable stationary zero state~$O$; two others 
correspond to some finite amplitudes. The fixed point~$A$ at larger amplitude 
is stable while that with less amplitude~$S$ is unstable. As value of~$h$ 
decreases, the fixed points~$A$ and~$S$ approach each other, merge 
(approximately at $h\approx -1.23$) and then disappear. Of course, this is 
only a preliminary rough picture, and we must discuss now what happens 
actually nearby this parameter value in the original system.

\begin{figure}
\includegraphics[width=0.45\textwidth,keepaspectratio]{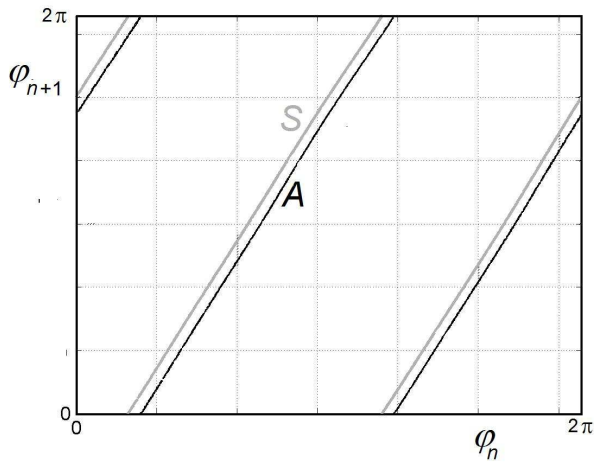}
\caption{Diagrams illustrating dynamics of phase on successive stages of 
excitation of the first oscillator on the attractor~$A$ (black) and on the 
non-attracting invariant set~$S$ (gray).}\label{f3}
\end{figure}

The fixed point~$A$ for the original system~\eqref{1} is associated with the 
Smale--Williams attractor: while the amplitude evolves almost periodically 
(with the period of the parameter modulation) the phases at successive periods 
behave in accordance with the chaotic Bernoulli map. Of the same kind is 
dynamics on a non-attractive invariant set corresponding to the unstable 
fixed point~$S$; the only difference is the additional instability of orbits 
in this set in respect to the amplitude variations, as evident from Fig.~\ref{f2}b. 
Henceforth, we use the designations~$A$ for the attractor and~$S$ for the 
non-attracting invariant set of the original system rather than for the fixed 
points of the amplitude map. Figure~\ref{f3} illustrates the phase dynamics on these 
sets with the computed iteration diagrams. The branches shown in black relate 
to the attractor, and the gray ones to the non-attractive invariant set. For 
the attractor the plot is obtained in straightforward way in the course of 
numerical integration of the equations~\eqref{1}~\cite{10,11,12,19}: we 
evaluate the phases~$\varphi_n=\arg[x_0(t_n)+ix_1(t_n)]$ from the instant 
states for the first oscillator at $t_n=nT$ and plot $\varphi_{n+1}$ versus 
$\varphi_n$. To do the same thing for the non-attractive invariant set~$S$ 
special measures are needed to keep the trajectory on it; actually, the picture 
was obtained from a collection of data for periodic orbits belonging to the set. 

Accounting that the dynamics for the phases on both invariant sets~$A$ and~$S$ 
correspond to topologically equivalent Bernoulli maps, the orbits belonging to 
each of them are in one-to-one correspondence with a set of infinite two-symbol 
(binary) sequences. 

It is well known that a chaotic attractor contains a dense set of unstable 
periodic orbits which constitutes a kind of "skeleton" for the attractor~\cite{22}. 
In terms of symbolic dynamics, it corresponds to a class of the orbits associated 
with periodic binary sequences. This "skeleton" is a convenient object to 
comprehend the dynamics on the whole attractor~$A$. The same is true for the 
non-attracting invariant set~$S$, which possesses its own "skeleton" of 
unstable periodic orbits. The both sets are in one-to-one correspondence 
(as follows from identical symbolic dynamics): each periodic orbit on the 
attractor has a partner ("dual orbit") belonging to the invariant set~$S$. 
Dealing with the "skeletons" it is possible to look over all periodic orbits 
(at least up to some reasonably large period) and analyze their bifurcations in 
dependence on the parameter~$h$.

A periodic orbit of the stroboscopic map belonging to the attractor~$A$ has a 
three-dimensional stable manifold and a one-dimensional unstable manifold (actually, 
the last one coincides with the whole attractor). The Floquet multiplier responsible 
for the instability may be evaluated as $\mu_1^A \approx 2^p$, where~$p$ is the 
period of the orbit. The next multiplier, associated with the amplitude 
perturbation~$\mu_2^A$ is less than $1$. (It may be estimated as $[f'(r_A)]^p$, 
where~$f$ is the function plotted in Fig.~\ref{f2}b, and~$r_A$ designates the position 
of the stable fixed point.) On the other hand, a periodic orbit belonging to the 
invariant set~$S$ possesses a two-dimensional stable manifold and a two-dimensional 
unstable manifold; one multiplier indicating instability in respect to the phase is 
$\mu_1^S \approx 2^p$, and another one, associated with the amplitude instability 
$\mu_2^S$, is essentially smaller: $1<\mu_2^S<\mu_1^S$. Under decrease of parameter~$h$, 
mutually dual orbits, one from the set~$A$ and another from the set~$S$ undergo a 
"fold" (or tangent, or saddle-node) bifurcation: they become closer, coalesce at 
some bifurcation value~$h_{\rm bif}$ and then disappear. 

If the description in terms of the amplitude map of Fig.~\ref{f2} be exact, the 
bifurcations would happen simultaneously for all pairs of the dual orbits, but 
actually it is not so: the bifurcation values~$h_{\rm bif}$ occupy a finite, 
although narrow interval. (Roughly, this interval $-1.22<h<1.24$ may be thought 
as associated with a finite width of the "curves" visible in the panel~(a) of~Fig.~\ref{f2}.) 

The drawn picture is supported by computations with Eqs.~\eqref{1}: we have traced 
the bifurcations for a large number of pairs of dual orbits associated with periodic 
symbolic codes. Table~1 collects the bifurcation values~$h_{\rm bif}$ for the orbits of 
period from $1$ to $5$ along with their binary codes.

According to the bifurcation theory~\cite{23}, the fold bifurcation may be regarded 
as corresponding to a \textit{turning point} of a curve, which represents a pair 
of periodic orbits in the extended space, which is the state space complemented 
with the coordinate axis of the parameter~$h$. This interpretation reflects the 
fact that for~$h>h_{\rm bif}$ each saddle cycle embedded in the attractor has, as a 
counterpart, a dual saddle cycle from the non-attracting invariant set. At the 
bifurcation point they both merge and annihilate, or, to put it another way, 
the respective branches of the curve continue each other.

Figure~\ref{f3}a plots coordinates of points of the periodic orbits of periods~$p\le10$ 
for the stroboscopic map of the system~\eqref{1} versus parameter~$h$ in the 
three-dimensional extended phase parameter space~$\{x_0,x_1,h\}$. On this picture 
attractor corresponds to the outer part of the tube-like formation, while the 
non-attracting invariant set corresponds to the inner part. It looks as going 
inside out in order to be eventually transformed into the Smale--Williams attractor 
after passage of the turning points. Figure~\ref{f3}b plots radial (amplitude) component 
for the same set of periodic orbits versus parameter~$h$; it may be regarded as a 
longitudinal slice of the object of panel~(a). 

\begin{table}
\caption{Bifurcation values of parameter~$h$ for the unstable periodic orbits 
of the system~\eqref{1} at $\omega_0=2\pi$, $A=6.5$, $T=6$, $\varepsilon=0.5$}

\begin{tabular}{|c|c|c|} 
\hline 
$p$ & Symbolic codes & $h_{\rm{bif}}$ \\ \hline 
$1$ & R & $-1.22583188$ \\ \hline 
$2$ & LR & $-1.23358210$ \\ \hline 
$3$ & LRR & $-1.22879290$ \\  
    & RLL & $-1.23489088$ \\ \hline 
$4$ & LRRR & $-1.22695959$ \\  
    & LRRL & $-1.23226821$ \\  
    & RLLL & $-1.23371422$ \\ \hline 
$5$ & LRRLR & $-1.23070842$ \\  
    & RRRRL & $-1.22628724$ \\  
    & LRRRL & $-1.23012599$ \\  
    & RLLLL & $-1.23249362$ \\  
    & LRRLL & $-1.23263428$ \\  
    & RLLRL & $-1.23467205$ \\ \hline 
\end{tabular}
\end{table}

\begin{figure}
\includegraphics[width=0.45\textwidth,keepaspectratio]{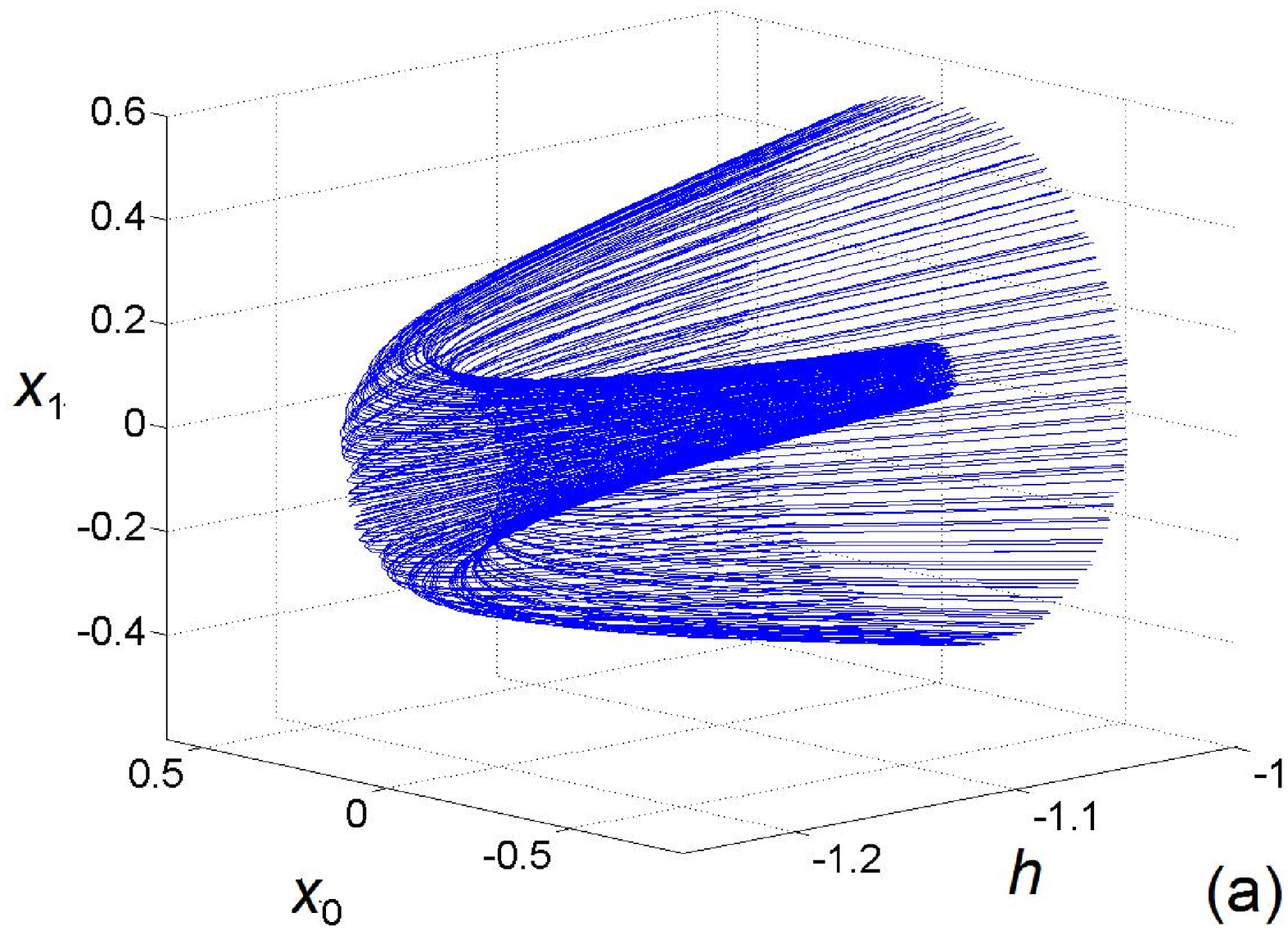}
\includegraphics[width=0.45\textwidth,keepaspectratio]{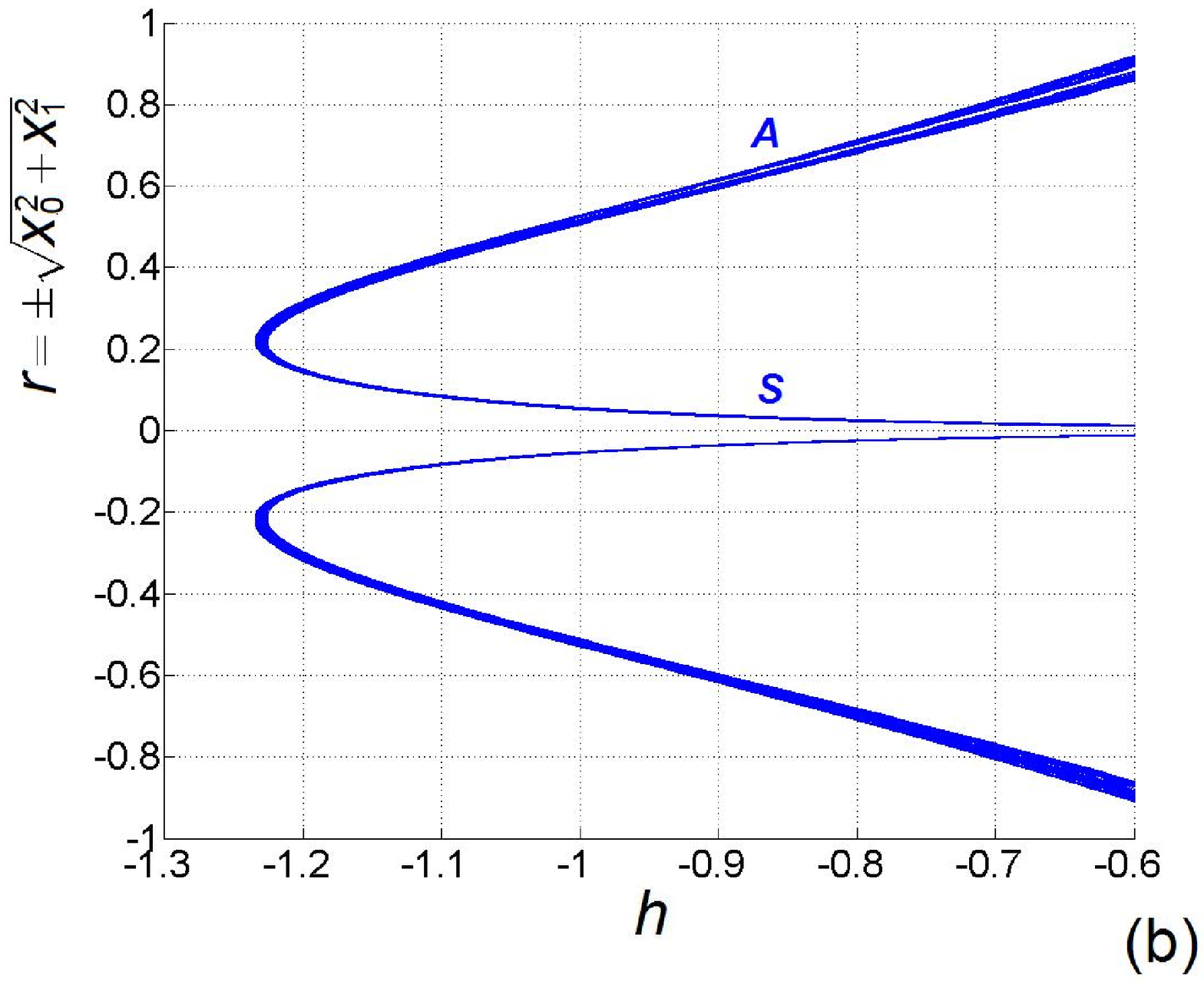}
\caption{(a) Periodic solutions of the equations~\eqref{1} with period $\le7$ 
plotted in the extended phase parameter space at $\omega_0=2\pi$, $A=6.5$, 
$T=6$, $\varepsilon=0.5$. (b) Superimposed graphs of radial (amplitude) component 
of the same set of solutions versus parameter~$h$; this is actually a longitudinal 
slice of the figure~(a).}\label{f4}
\end{figure}

\begin{figure}
\includegraphics[width=0.45\textwidth,keepaspectratio]{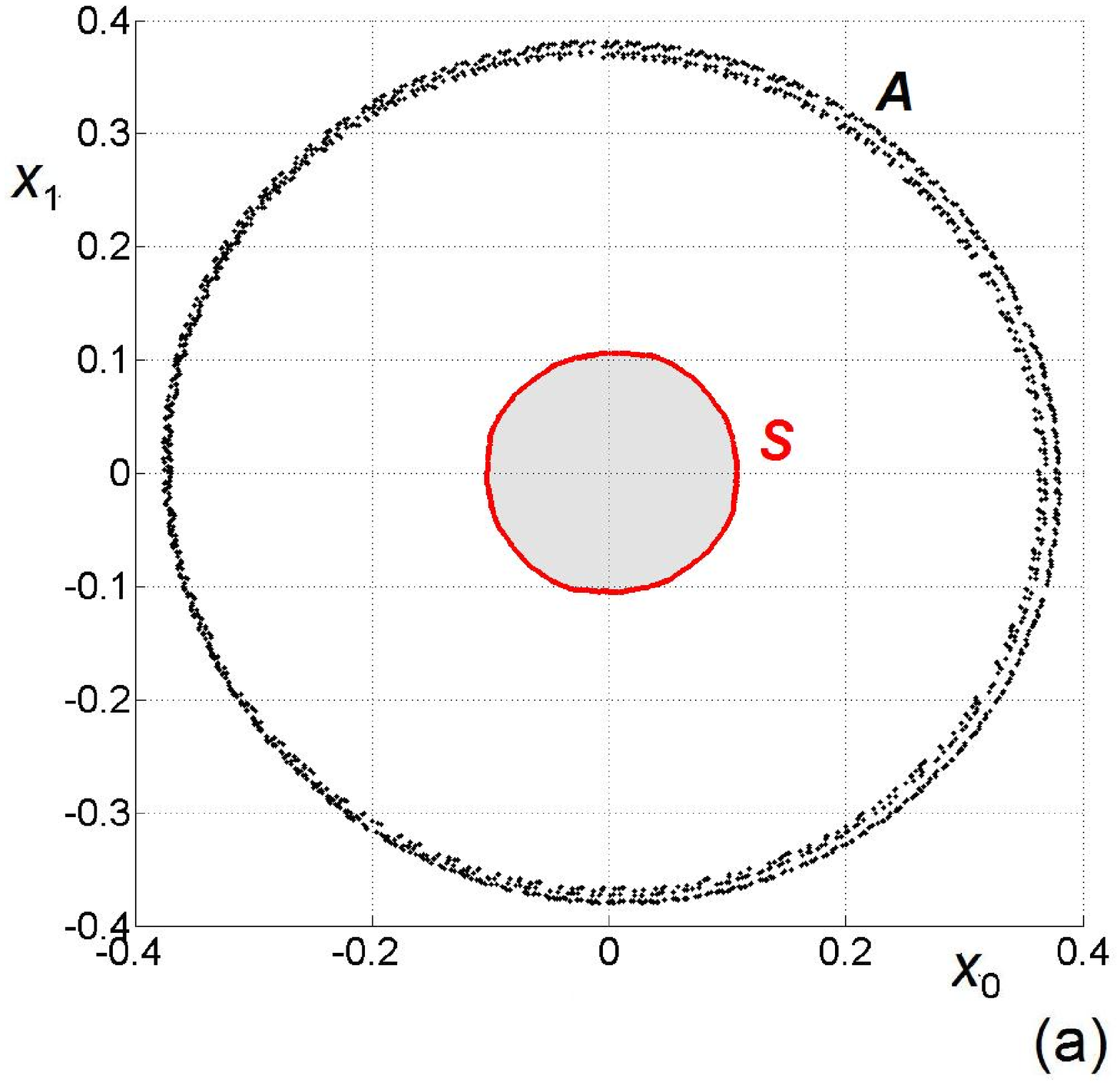}
\includegraphics[width=0.45\textwidth,keepaspectratio]{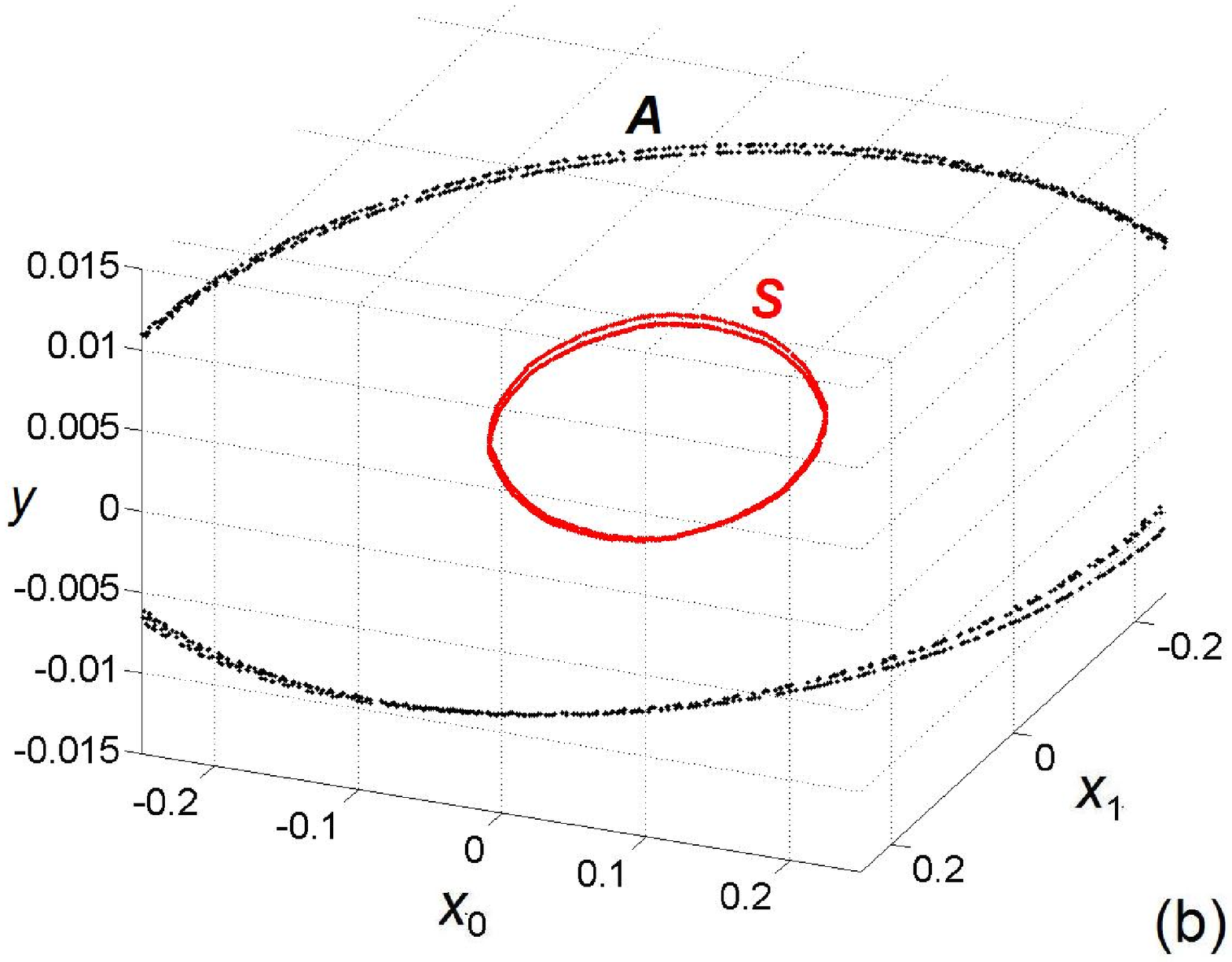}
\caption{Attractor~$A$ and non-attracting invariant set~$S$ in the stroboscopic 
Poincar\'{e} section shown on the plane of variables of the first oscillator~(a) 
and in three-dimensional plot~(b) accounting a variable of the second oscillator 
at $\omega_0=2\pi$, $A=6.5$, $T=6$, $\varepsilon=0.5$, and $h=-1.15$. The gray 
color on the panel~(a) indicates roughly the basin of attraction of the stable 
fixed point in the origin.}\label{f5}
\end{figure}

The structure observed at fixed $h=-1.15$ is illustrated in Fig.~\ref{f5}. Panel~(a) represents, 
in fact, a transversal section of the object shown in Fig.~\ref{f4}a. Observe two formations 
looking in rough approximation as closed circular curves but actually having fine 
transversal fractal structure. The outer one represents the chaotic attractor~$A$, 
while the inner one is the non-attracting invariant set~$S$ consisting of saddle 
orbits dual to those embedded in the attractor. It should be remembered, that the 
picture actually is a projection of the corresponding sets from the four-dimensional 
state space of the stroboscopic Poincar\'{e} map onto the plane. In Fig.~\ref{f5}b the 
three-dimensional version of the diagram is shown, and one may see more details of 
the fractal-like transversal structure of both sets~$A$ and~$S$. Evidently, the 
invariant set~$S$ and its stable invariant manifold separate the basin of attraction 
of the Smale--Williams solenoid and that of the attractive fixed point in the origin. 
(In Fig.~\ref{f5}a the last one corresponds roughly to gray area inside the inner 
closed "curve".)
 
Objects similar to the invariant set~$S$ are encountered in the theory of complex 
maps being known as Julia sets~\cite{24,25}. This analogy is rather deep: far from 
the bifurcation transition, when the size of the set~$S$ becomes small enough 
one can neglect all nonlinearities in the equations, beside that in the coupling 
term. Then, in a frame of description of the oscillators in terms of complex 
amplitudes, this set really turns to a Julia set for some complex map~\cite{26}.

Qualitatively, the content of the considered scenario is illustrated by a simple 
artificially constructed model map in Appendix (in a similar way like one-dimensional 
maps illustrate the Feigenbaum or intermittent scenarios).

\section{Conclusion}

Understanding scenarios for appearance of uniformly hyperbolic attractors 
under variation of control parameters seems relevant for a search for 
real-world systems with such attractors, which will be of special 
practical and theoretical interest because of structural stability of 
the generated chaos. In this article, basing on numerical computations, 
we reveal a nature of the bifurcation scenario of birth or destruction 
of a uniformly hyperbolic attractor of Smale--Williams type in stroboscopic 
Poincar\'{e} map of two alternately excited non-autonomous van der Pol oscillators 
under variation of a parameter controlling relative duration of stages of activity 
and silence of the oscillators. This is the second discussed in the literature 
scenario relating to formation of a uniformly hyperbolic attractor, after that of
 Shil'nikov and Turaev~\cite{16,17,18} corresponding to a kind of the so-called 
blue-sky catastrophe. In our scenario, the birth or destruction of the 
Smale--Williams solenoid appears not as a single bifurcation event, but 
occupies a finite parameter interval; it may be thought as a multitude of 
saddle-node bifurcations each involving a pair of orbits, one from the 
attractor, and another from the non-attracting invariant set.

\section*{Acknowledgement}
The work was supported, in part, by RFBR-DFG grant No 11-02-91334. 
O.I. acknowledges support from Grant of President of Russian Federation 
MK-905.2010.2 and RFBR grant No 12-02-00342.

\section*{Appendix}

\begin{figure}
\includegraphics[width=0.45\textwidth,keepaspectratio]{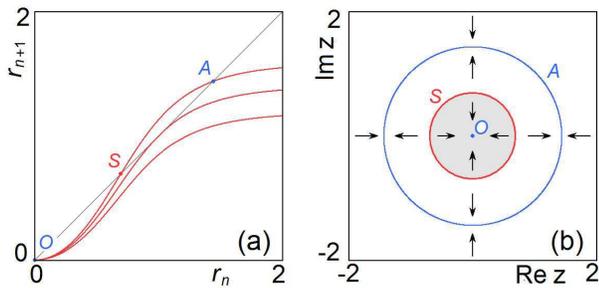}
\caption{A plot of radial component of the model map~\eqref{2}~(a) and disposition 
of attractors~$A$, $O$ and of a non-attractive invariant set~$S$ on the plane of 
complex variable~$z$~(b).}\label{fa1}
\end{figure}

It is a good way to illustrate the considered scenario with a simple model. 
Let us start with the equation:
\begin{equation} \label{2} 
z_{n+1}=\frac{Rz_n^2}{\sqrt{1+|z_n|^4}},
\end{equation} 
where~$z$ is a complex variable, and~$R$ is a real parameter. Setting 
$z=re^{i\varphi }$ we rewrite the map as
\begin{equation} \label{3} 
r_{n+1}=R\frac{r_n^2}{\sqrt{1+r_n^4}}, \qquad 
\varphi_{n+1}=2\varphi_n \quad (\rm{mod} 2\pi).
\end{equation} 
At small~$R$ the map for the radial variable~$r$ has only a single fixed point 
at zero. With increase of~$R$ at some instant ($R=\sqrt{2}$), a tangent 
bifurcation occurs accompanied with appearance of a pair of fixed points 
$r_{1,2}=\sqrt{{\tfrac{1}{2}} R^2\mp\sqrt{{\tfrac{1}{4}} R^4-1} }$, one stable 
and another unstable (see Fig.~\ref{fa1}). For the complex map~\eqref{1} the 
stable fixed point corresponds to attractor~$A$ placed on a circle of radius~$r_2$ 
while the angular coordinate is governed by the Bernoulli map 
$\varphi_{n+1}=2\varphi_n\quad (\rm{mod}2\pi)$. The unstable point corresponds 
to an unstable invariant curve that is a circle of radius $r_1$ on which the 
angular variable evolves in accordance with the same Bernoulli map, and this 
curve is a boundary separating basins of two coexisting attractors, the stable 
point~$O$ and the attractor~$A$. 

Now, consider a slightly modified version of the map 
\begin{equation} \label{4} 
z_{n+1}=\frac{Rz_n(z_n+\varepsilon)}{\sqrt{1+|z_n (z_n+\varepsilon)|^2}},
\end{equation} 
which reduces to~\eqref{2} at $\varepsilon=0$, and look what happens under 
increase of parameter~$R$ with $\varepsilon\ne 0$. At small~$R$ the unique 
attractor is a zero fixed point~$O$. For~$R$ great enough there is another 
attractor~$A$, situated in the region of relatively large~$|z|$. Concerning 
dynamics on this attractor, the additional terms in the map~\eqref{4} do not 
violate the nature of the dynamics of the angular variable, which follows a map 
of the same topological type as the Bernoulli map. However, now the invariant 
set~$A$ does not coincide precisely with the circle, but acquires a transversal 
split and looks similar to a projection of the Smale--Williams solenoid. The border 
between the basins of attractors~$O$ and~$A$ is represented now by some non-trivial 
invariant set~$S$, which may be computed using backward iterations of the map~\eqref{4}:
\begin{equation} \label{5} 
z_n=-\frac{\varepsilon}{2} \pm \sqrt{\frac{\varepsilon^2}{4}+
\frac{z_{n+1}}{\sqrt{R^2-|z_{n+1}|^2}}},
\end{equation} 
where a sign at the square root is selected on each next iteration randomly. 
Visually, the set~$S$ looks similar to a Julia set of complex quadratic map. 
(It is not surprising: neglecting the nonlinear term in the denominator of~\eqref{4} 
we get just the complex analytic map $z_{n+1}=R(z_n^2+\varepsilon z_n)$ providing a reasonable 
approximation in the domain of location of the set~$S$ while~$|z|$ is small.) 
Dynamics on the sets~$A$ and~$S$ may be described with the two-symbol Bernoulli 
scheme. (For the set~$S$ it is obvious from the above algorithm of backward iterations: 
a symbolic sequence of an orbit corresponds to the sequence of signs selected in the 
course of the procedure.) 

\begin{figure*}
\includegraphics[width=0.9\textwidth,keepaspectratio]{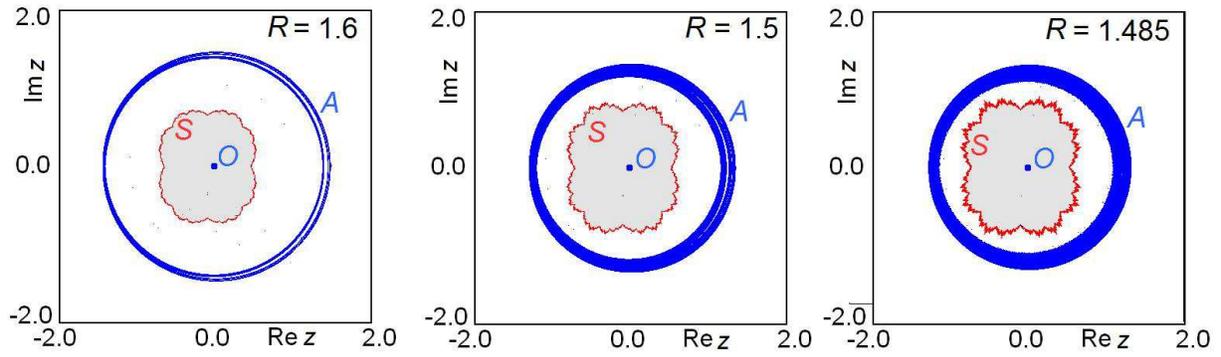}
\caption{Mutual disposition of attractors~$A$ and~$O$ and of a non-attractive 
invariant set ($S$) of the map~\eqref{4} on the plane of complex variable~$z$ 
for different values of parameter~$R$ at~$\varepsilon=0.2$.}\label{fa2}
\end{figure*}

With decrease of parameter~$R$ a size of the attractor~$A$ reduces while a size 
of the set~$S$ grows (Fig.~\ref{fa2}). As the sets~$A$ and~$S$ approach each 
other, the transverse structure of~$S$ develops in the radial direction, and 
becomes less akin to the original Julia set. For some~$R=R_1$ a touch of~$A$ 
and~$S$ happens, and it corresponds to the moment of crisis of the attractor~$A$ 
(the touch of the basin boundary, cf.~\cite{21}). 

Each of the sets~$A$ and~$S$ contains a dense set of unstable periodic orbits. 
A periodic orbit on the attractor~$A$ has a multiplier~$\mu_1^A\cong 2^p>1$, 
where~$p$ is a period of the orbit, and a multiplier $\mu_2^A<1$. For an orbit 
on the invariant set~$S$ the multipliers are $\mu_1^S\cong 2^p>1$ and $\mu_2^S>1$. 
As symbolic representations of the orbits in~$A$ and~$S$ are of the same type, 
each orbit in~$A$ can be related to a partner dual orbit in~$S$ and vice versa. 
With decreasing~$R$ the dual orbits approach each other, then merge and disappear 
via the fold (tangent) bifurcation, where $\mu_2^A=\mu_2^S=1$. For different 
periodic orbits this bifurcation occurs at different values of~$R$, so that the 
process of disappearance of orbits remaining on the set~$A$ occupies a finite 
parameter interval~$[R_2, R_1]$. Only in the degenerate case $\varepsilon=0$, 
all orbits belonging to the sets~$A$ and~$S$ undergo the bifurcations 
simultaneously, just at~$R=\sqrt{2}$. 

The above consideration is in clear qualitative correspondence with that in the main 
part of the article. Some inconsistence is that the model map~\eqref{4} is 
non-invertible (in contrast to the Poincar\'{e} map of the system~\eqref{1}). 
This issue could be repaired with a H\'{e}non-like modification of the model: 
$z_{n+1}=\frac{Rz_n(z_n+\varepsilon)}{\sqrt{1+|z_n(z_n+\varepsilon)|^2}}-bz_{n-1}$, 
but in this case the analysis and explanations become much less transparent. (We 
can keep up appearances saying that we deal with this invertible map, but assign 
exclusively small value to the parameter~$b$.)


\begin {thebibliography}{99}

\bibitem{1} A.~Katok and B.~Hasselblatt, \textit{Introduction to the 
Modern Theory of Dynamical Systems} (Cambridge University Press, New York, 1995).

\bibitem{2} L.~Shilnikov,  Int. J. of Bifurcation and Chaos 
\textbf{7}, 1353 (1997).

\bibitem{3} D.V.~Anosov, G.G.~Gould, S.K.~Aranson et al, 
 in \textit{Encyclopaedia of Mathematical Sciences}, Vol.~9 
(Springer, Berlin, 1995).

\bibitem{4} S.~Smale, Bull. Amer. Math. Soc. (NS) \textbf{73}, 747 (1967).

\bibitem{5} R.F.~Williams, Publications math\'{e}matiques de 
l'I.H.\'{E}.S. \textbf{43}, 169 (1974).

\bibitem{6} R.V.~Plykin, Math. USSR Sb. \textbf{23}(2), 233 (1974). 
(In Russian)

\bibitem{7} C.~Bonatti, L.J.~Diaz, M.~Viana, Dynamics Beyond 
Uniform Hyperbolicity. A Global Geometric and Probobalistic Perspective. 
\textit{Encyclopedia of Mathematical Sciences}, Vol.102, (Springer, 
Berlin, 2005).

\bibitem{8} V.S.~Anishchenko, \textit{Nonlinear dynamics of chaotic 
and stochastic systems: tutorial and modern developments} (Springer, Berlin, 2002).

\bibitem{9} A.~Loskutov, Physics-Uspekhi \textbf{53}, 1257 (2010).

\bibitem{10}S.P.~Kuznetsov: Phys. Rev. Lett. \textbf{95}, 144101 (2005).

\bibitem{11} S.P.~Kuznetsov and A.~Pikovsky, Physica D \textbf{232}, 87 (2007).

\bibitem{12} S.P.~Kuznetsov. \textit{Hyperbolic Chaos: A Physicist's 
View. Series: Mathematical Methods and Modeling for Complex Phenomena} 
(Higher Education Press, Beijing and Springer, Berlin, 2012).

\bibitem{13} S.P.~Kuznetsov. Physics-Uspekhi \textbf{54} (A.1), 119 (2011).

\bibitem{14} D.~Ruelle and F.~Takens, Commun. Math. Phys. 
\textbf{20}, 167 (1971).

\bibitem{15} S.~Newhouse, D.~Ruelle, and F.~Takens. Commun. Math. Phys. 
\textbf{64}, 35 (1978).

\bibitem{16} L.P.~Shil'nikov and D.V.~Turaev. Doklady Akademii Nauk 
\textbf{342}, no.~5, 596 (1995).

\bibitem{17}  L.P.~Shil'nikov and D.V.~Turaev. Computers Math. Appl. 
\textbf{34}, 173 (1997).

\bibitem{18} S.P.~Kuznetsov. Regular and Chaotic Dynamics 
\textbf{15}, No. 2-3, 348 (2010). 

\bibitem{19} S.P.~Kuznetsov and I.R.~Sataev, Physics Letters 
A \textbf{365}, 97 (2007). 

\bibitem{20} D.~Wilczak, SIAM J. Applied Dynamical Systems 
\textbf{9}, 1263 (2010).

\bibitem{21} C.~Grebogi, E.~Ott, J.A.~Yorke. Physica D 
\textbf{7}, 181 (1983).

\bibitem{22} P.~Cvitanovi\'{c}. Physica D, 
\textbf{51}, 138 (1991).

\bibitem{23} Yu.A.~Kuznetsov. \textit{Elements of Applied Bifurcation 
Theory} (Springer, Berlin, 1998).

\bibitem{24} R.L.~Devaney. \textit{An Introduction to Chaotic Dynamical 
Systems} (Westview Press, New York, Addison-Wesley 2003).

\bibitem{25} O.~Biham and W.~Wenzel. Phys. Rev. A \textbf{42}, 4639 (1990).

\bibitem{26} O.B.~Isaeva, S.P.~Kuznetsov and A.H.~Osbaldestin. Physica D 
\textbf{237}, 873 (2008).

\end{thebibliography}
\end{document}